\documentclass[pra,amsmath]{revtex4}
\usepackage{pifont}
\usepackage{graphicx,epsfig,subfigure,dsfont,amssymb,amsmath,amsthm,amsfonts,amsbsy,mathrsfs,amscd}

\usepackage[all]{xy}

\theoremstyle{definition}

\begin{document}

\title{\bf Dynamics of coherence-induced state ordering under Markovian channels}
\author{Long-Mei Yang$^1$, Bin Chen$^2$\thanks{Corresponding author: chenbin5134@163.com.}, Shao-Ming Fei $^1$, Zhi-Xi Wang$^1$
\\
{\footnotesize $^1$School of Mathematical Sciences, Capital Normal University, Beijing 100048, China}\\
{\footnotesize $^2$School of Mathematical Sciences, Tianjin Normal University, Tianjin 300387, China}}

\begin{abstract}
We study the dynamics of coherence-induced state ordering under incoherent channels, particularly
 four specific Markovian channels:
$-$ amplitude damping channel, phase damping channel, depolarizing channel and bit flit channel for single-qubit states.
We show that the amplitude damping channel, phase damping channel, and depolarizing channel
do not change the coherence-induced state ordering
by $l_1$ norm of coherence, relative entropy of coherence, geometric measure of coherence, and Tsallis relative $\alpha$-entropies, while
the bit flit channel does change for some special cases.
\end{abstract}

\maketitle


\section{Introduction}
Quantum coherence is a fundamental feature of quantum mechanics, which distinguishes the quantum world from the classical physics realm.
It is an important aspect in many research fields such as low-temperature thermodynamics \cite{berg1,nara,cwik,lost1,lost2},
quantum biology \cite{plenio,reben,lloyd,li,huel,levi}, and nanoscale physics \cite{vazq,karl}.
Quantifying the coherence of quantum states \cite{berg2} has become a topic of interest for researchers.
Baumgratz {\it et al.} have recently proposed a strict framework to quantify quantum coherence \cite{baum}.
Various coherence measures have been defined based on this framework,
such as $l_1$-norm of coherence, relative entropy of coherence \cite{baum}, geometric measure of coherence \cite{stre}
and Tsallis relative $\alpha$-entropies of coherence measure \cite{zhang}.
Here, the Tsallis relative $\alpha$-entropy of coherence measure violates the condition of a coherence measure that
does not increase under mixing of states, while
it satisfies a generalized monotonicity of average coherence under subselection based on measurement.

Different coherence measures employed based on different physical contexts thus give rise to different values of coherence.
Questions about ordering states with various coherence measures have also been discussed \cite{zhang,coherent,liu}.
Whether or not the quantum operators change the coherence-induced state ordering proposed by Zhang {\it et al} \cite{coherent}, forms another interesting problem.

Focused on single-qubit states, in this study, we investigate such ordering problems under incoherent channels. In particular, we consider four Markovian channels: $-$ amplitude damping channel, phase damping channel, depolarizing channel, and bit flit channel.
Note that for some special cases, Zhang {\it et al} have already studied the problem for single-qubit states by using the amplitude damping channel and phase damping channel \cite{coherent}.
Here, we also consider the geometric measure of coherence for more general situations.
We extend the results of Ref. \cite{coherent} to general cases.
Furthermore, we show that the depolarizing channel does not change the coherence-induced state ordering
while the bit flit channel changes it when $p=\frac{1}{2}$.

\section{Preliminaries}
In this section, we first recapitulate some concepts related to quantum coherence.
Let $\mathcal{H}$ be a $d$-dimensional Hilbert space and $\{|i\rangle\}_{i=0}^{d-1}$ be an orthonormal basis of $\mathcal{H}$.
An incoherent state is defined as $\rho=\Sigma_{i=0}^{d-1}p_i|i\rangle\langle i|$, where $p_i\geq0,\Sigma_ip_i=1$.
Let $\mathcal{I}$ denote the set of incoherent states.
An incoherent operation is defined as $\Lambda(\rho)=\Sigma_nK_n\rho K_n^{\dag}$,
where $\Sigma_nK_nK_n^{\dag}=I$ and $K_n\mathcal{I}K_n^{\dag}\subset\mathcal{I}$.
Baumgratz {\it et al.} proposed a framework to quantify quantum coherence, that is,
a function $\mathcal{C}$ can be taken as a coherence measure if it satisfies the following postulates \cite{baum}:

{\rm (C1)} $\mathcal{C}(\rho)\geq0,\ \mathcal{C}(\rho)=0$ if and only if $\rho\in\mathcal{I}$;

{\rm (C2)} $\mathcal{C}(\Lambda(\rho))\leq\mathcal{C}(\rho)$ for any incoherent operation $\Lambda$;

{\rm (C3)} $\Sigma_np_n\mathcal{C}(\rho_n)\leq \mathcal{C}(\rho)$, where $p_n={\rm Tr}(K_n\rho K_n^{\dag})$ and $\rho_n=K_n\rho K_n^{\dag}/p_n$,
$\{K_n\}$ is a set of incoherent Kraus operators;

{\rm (C4)} $\mathcal{C}(\Sigma _i p_i\rho_i)\leq\Sigma _i p_i\mathcal{C}(\rho_i)$ for any
set of quantum states $\{\rho_i\}$ and any probability distribution $\{p_i\}$.

Several coherence measures have been put forward based on this framework.
Here, we give the definitions of the following four coherence measures for further use.

Let $\rho$ be a state defined on $\mathcal{H}$, then
\begin{equation}\label{l1}
\mathcal{C}_{l_1}(\rho)=\sum\limits_{i\neq j}|\rho_{ij}|
\end{equation}
is the $l_1$ norm of coherence,
where $\rho_{ij}$ are the entries of $\rho$.
The relative entropy of coherence is defined by
\begin{equation}\label{rel}
\mathcal{C}_r(\rho)=\underset{\sigma\in\mathcal{I}}{\min}\mathcal{S}(\rho\|\sigma)=\mathcal{S}(\rho_{diag})-\mathcal{S}(\rho),
\end{equation}
where $\mathcal{S}(\rho\|\sigma)={\rm Tr}(\rho\log\rho-\rho\log\sigma)$ is the quantum relative entropy,
$\mathcal{S}(\rho)=- {\rm Tr}(\rho\log\rho)$ is the von Neumann entropy, and $\rho_{diag}=\Sigma_i\rho_{ii}|i\rangle\langle i|$
is the diagonal part of $\rho$.
The geometric measure of coherence is defined by
\begin{equation}\label{gem}
\mathcal{C}_g(\rho)=1-\underset{\sigma\in\mathcal{I}}{\max}F(\rho,\sigma),
\end{equation}
where $F(\rho,\sigma)=\Big({\rm Tr}\sqrt{\sqrt{\sigma}\rho\sqrt{\sigma}}\Big)^2$ is the fidelity of two density operators $\rho$
and $\sigma$.
The Tsallis relative $\alpha$-entropy of coherence is defined by
\begin{equation}\label{Tsa}
\mathcal{C}_\alpha(\rho)=\underset{\delta\in\mathcal{I}}{\min}\mathcal{D}_\alpha(\rho\|\delta)=
\frac{r^\alpha-1}{\alpha-1},
\end{equation}
where $r=\Sigma_i\langle i|\rho^\alpha|i\rangle^{\frac{1}{\alpha}}$ and $\alpha\in(0,1)\cup(1,2]$.

Any single-qubit state can be expressed as
\begin{equation}\label{sin}
\rho=\frac{1}{2}(I+\vec{k}\vec{\sigma})=\frac{1}{2}(I+t\vec{n}\vec{\sigma}),
\end{equation}
where $\vec{k}=(k_x,k_y,k_z)$ is a real vector satisfying $\|\vec{k}\|\leq1$, $t=\|\vec{k}\|$,
$\vec{n}=(n_x,n_y,n_z)$ is a unit vector, and $\vec{\sigma}=(\sigma_x,\sigma_y,\sigma_z)$ is the vector of Pauli matrices.
Here, we note that $n_x, \ n_y, \ n_z$ represent the length of vector $\vec{k}$ along the direction $\sigma_x,\sigma_y,\sigma_z$,
respectively.

A non-coherence-generating channel (NC) $\tilde{\Phi}$  is a completely positive and trace preserving (CPTP) map from an incoherent state to another incoherent state:
$\tilde{\Phi}(\mathcal{I})\subset \mathcal{I}$, where $\mathcal{I}$ denotes the set of incoherent states \cite{NC}.
Any quantum channel $\Phi$ is called an incoherent channel if there exists a Kraus decomposition
$\Phi(\cdot)=\Sigma_nK_n(\cdot)K_n^{\dagger}$ such that $\rho_n=\frac{K_n(\rho)K_n^{\dagger}}{{\rm Tr}(K_n(\rho)K_n^{\dagger})}$
is incoherent for any incoherent state $\rho$.

A rank-$2$ qubit channel is an NC if and only if it has the Kraus decomposition either as \cite{NC}
\begin{equation}\label{eqnc1}
\Phi^{(1)}(\cdot)=E_1^{(1)}(\cdot)E_1^{(1)\dagger}+E_2^{(1)}(\cdot)E_2^{(1)\dagger}
\end{equation}
with
\begin{equation}\label{eqNC1}
E_1^{(1)}=\left(
        \begin{array}{cc}
          e^{i\eta}\cos\theta\cos\phi & 0 \\
          -\sin\theta\sin\phi & e^{i\xi}\cos\phi \\
        \end{array}
      \right), \ \ \
     E_2^{(1)}=\left(
\begin{array}{cc}
                  \sin\theta\cos\phi & e^{i\xi}\sin\phi \\
                  e^{-i\eta}\cos\theta\sin\phi & 0 \\
                \end{array}
              \right)
\end{equation}
or as
\begin{equation}\label{eqnc2}
\Phi^{(2)}(\cdot)=E_1^{(2)}(\cdot)E_1^{(2)\dagger}+E_2^{(2)}(\cdot)E_2^{(2)\dagger}
\end{equation}
 with
\begin{equation}\label{eqNC2}
E_1^{(2)}=\left(
        \begin{array}{cc}
         \cos\theta & 0 \\
         0 & e^{i\xi}\cos\phi \\
        \end{array}
      \right),  \ \ \
      E_2^{(2)}=\left(
                \begin{array}{cc}
                  0 & \sin\phi \\
                  e^{i\xi}\sin\theta & 0 \\
                \end{array}
              \right),
\end{equation}
where $\theta,\phi,\xi$, and $\eta$ are all real numbers.
Here $\Phi^{(1)}$ is not an incoherent channel unless
$\sin\theta\cos\theta\sin\phi\cos\phi=0$ and $\Phi^{(2)}$ is an incoherent channel.

\section{Main results}
In this section, we first study the coherence-induced ordering problem
under arbitrary incoherent channels for single-qubit states via the coherence measures
$\mathcal{C}_{l_1}$, $\mathcal{C}_r$, $\mathcal{C}_\alpha$, and $\mathcal{C}_g$.
Then we study the dynamics of coherence-induced state ordering under specific Markovian channels for single-qubit states
by four Markovian channels namely, amplitude damping, phase damping channel, depolarizing channel, and bit flit channel.

Suppose that an incoherent channel is defined as in Eq. \eqref{eqnc1}.
Let $a=\frac{1-tn_z}{2}$ and $b=\frac{t(n_x-in_y)}{2}$ with $b=\mid b\mid e^{i\beta}$.
Then $\Phi(\rho)=\left(
                       \begin{array}{cc}
                         A & B\\
                         B^* & 1-A\\
                       \end{array}
                     \right)$
with $A=a\cos^2\phi+(b^*e^{i\xi}+be^{-i\xi})\sin\theta\sin\phi \cos\phi+(1-a)\sin^2\phi$,
$B=be^{i\eta-i\xi}\cos\theta\cos^2\phi+b^*e^{i\xi+i\eta}\cos\theta\sin^2\phi$.
Thus, $\mathcal{C}_{l_1}(\Phi(\rho))=2\mid be^{i\eta-i\xi}\cos\theta
\cos^2\phi+b^*e^{i\xi+i\eta}\cos\theta\sin^2\phi\mid$.
If $\sin\theta=0$, then $\Phi$ is an incoherent operation and $\mathcal{C}_{l_1}(\Phi(\rho))=2\mid b\mid\sqrt{e^{i\beta-i\xi}\cos^2\phi+e^{i\xi-i\beta}\sin^2\phi}$.
We find that the value of $\mathcal{C}_{l_1}(\rho)$ depends on both $b$ and the channel itself.
In other words, there may exist incoherent channels such that
$\mathcal{C}_{l_1}(\Phi(\rho_1))<\mathcal{C}_{l_1}(\Phi(\rho_2))$ though $\mathcal{C}_{l_1}(\rho_1)>\mathcal{C}_{l_1}(\rho_2)$.

Suppose that an incoherent channel is defined as in Eq. \eqref{eqnc2}.
Then $\Phi(\rho)=\left(
                       \begin{array}{cc}
                         C & D\\
                         D^* & 1-C\\
                       \end{array}
                     \right)$
with $C=a\cos^2\theta+(1-a)\sin^2\phi$ and $D=e^{i\xi}(b\cos\theta\cos\phi+b^*\sin\theta\sin\phi)$.
Thus, $\mathcal{C}_{l_1}(\rho)=2|b|\sqrt{\cos^2\beta\cos^2(\theta-\phi)+\sin^2\beta\cos^2(\theta+\phi)}$.
Also, we know that the value of $\mathcal{C}_{l_1}(\rho)$ depends on both $b$ and the channel itself.
In other words, there may exist incoherent channels such that
$\mathcal{C}_{l_1}(\Phi(\rho_1))<\mathcal{C}_{l_1}(\Phi(\rho_2))$ though $\mathcal{C}_{l_1}(\rho_1)>\mathcal{C}_{l_1}(\rho_2)$.

According to the above discussion, we can conclude that there exist incoherent
channels changing the coherence-induced state ordering under the coherence measure $\mathcal{C}_{l_1}$.
This is true also for the coherence measure $\mathcal{C}_g$, since $\mathcal{C}_{l_1}$ and $\mathcal{C}_g$ give the same ordering for single-qubit states \cite{ziji}.
For the other coherence measures $\mathcal{C}_r$ and $\mathcal{C}_\alpha$,
the issue will become formidably difficult for general incoherent channels.
However, we can consider some specific incoherent channels to deal with the problem.

\subsection{Amplitude damping channel}

The amplitude damping channel is characterized by the Kraus' operators:
$K_0=|0\rangle\langle0|+\sqrt{p}|1\rangle\langle1|, \ K_1=\sqrt{1-p}|0\rangle\langle1|$, where $p\in[0,1]$.
It can be directly verified that \cite{coherent},
\begin{equation}
\varepsilon (\rho)=\left(
                     \begin{array}{cc}
                       \frac{1+tn_z}{2}+\frac{p(1-tn_z)}{2} & \frac{\sqrt{1-p}t(n_x-\mathbf{i}n_y)}{2} \\
                       \frac{\sqrt{1-p}t(n_x+\mathbf{i}n_y)}{2} & \frac{(1-p)(1-tn_z)}{2} \\
                     \end{array}
                   \right),
\end{equation}
\begin{equation}
\mathcal{C}_{l_1}(\varepsilon(\rho))=(1-p)t\sqrt{1-n_z^2},
\end{equation}
\begin{equation}
\mathcal{C}_r(\varepsilon(\rho))=h\Big(\frac{1+t^{\prime}n_z^{\prime}}{2}\Big)-h\Big(\frac{1+t^{\prime}}{2}\Big),
\end{equation}
\begin{equation}
\mathcal{C}_{\alpha}(\varepsilon(\rho)=\frac{r^{\alpha}-1}{\alpha-1},
\end{equation}
where
$h(x)=-x\log x-(1-x)\log (1-x)$,
$r=\Big[\Big(\frac{1+t^{\prime}}{2}\Big)^{\alpha}\frac{1+n_z^{\prime}}{2}+
(\frac{1-t^{\prime}}{2}\Big)^{\alpha}\frac{1-n_z^{\prime}}{2}\Big]^{\frac{1}{\alpha}}
+\Big[\Big(\frac{1+t^{\prime}}{2}\Big)^{\alpha}\frac{1-n_z^{\prime}}{2}
+(\frac{1-t^{\prime}}{2}\Big)^{\alpha}\frac{1+n_z^{\prime}}{2}\Big]^{\frac{1}{\alpha}}$,
$t^{\prime}=\sqrt{(1-p)t^2(1-n_z^2)+(p+(1-p)n_zt)^2}$, $n_x^{\prime}=\frac{\sqrt{1-p}n_xt}{t^{\prime}}$,
$n_y^{\prime}=\frac{\sqrt{1-p}n_yt}{t^{\prime}}$, and $n_z^{\prime}=\frac{p+(1-p)n_zt}{t^{\prime}}$.

It is clear that
for the case $p=1$, the amplitude damping channel transforms any single-qubit state to an incoherent state.
For $p=0$, any single-qubit state is unchanged under amplitude damping channel.

\begin{figure}
\centering
\subfigure[~$p=\frac{1}{4}$]{
\label{fig:subfig:a}
\includegraphics[width=1.6in]{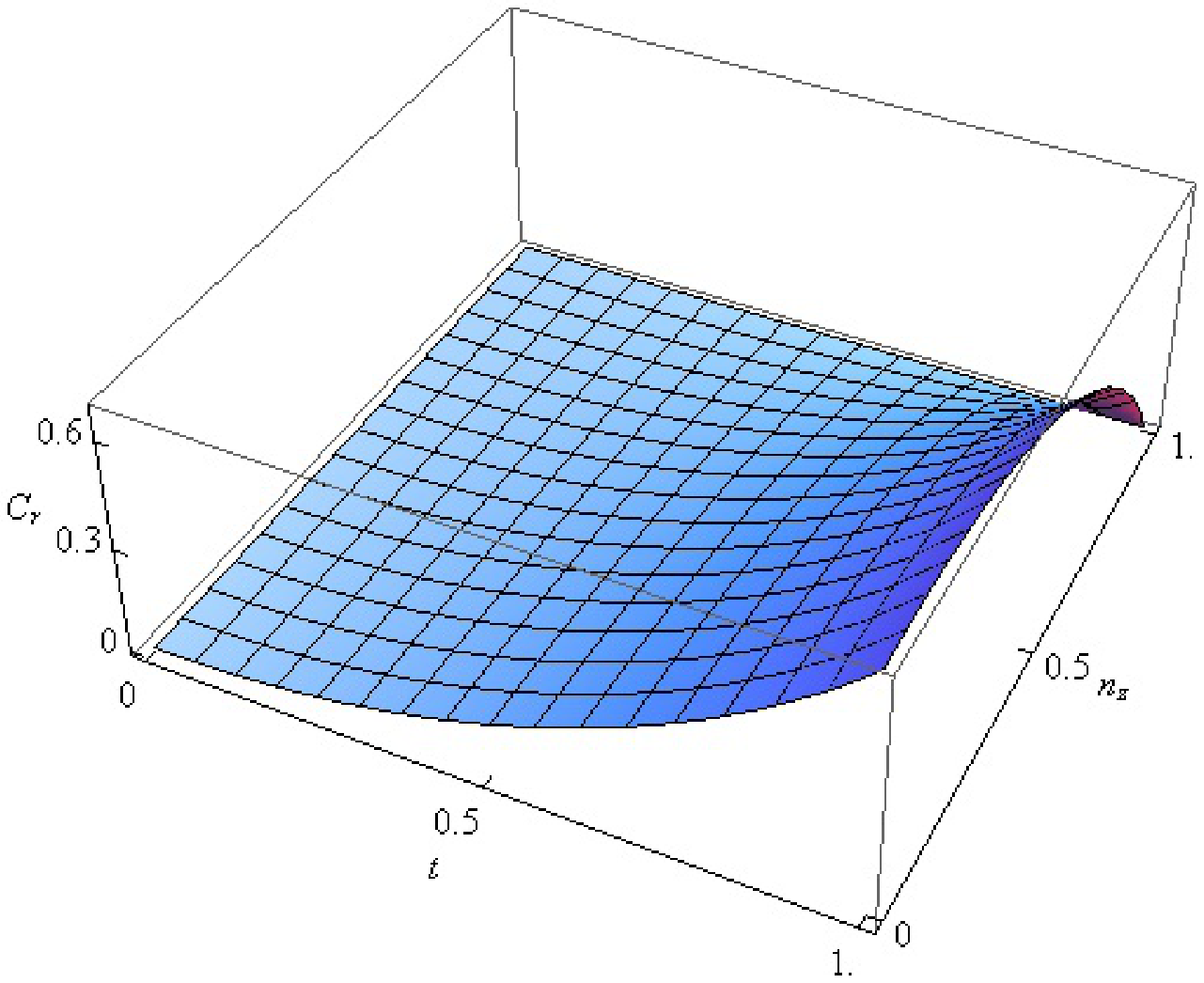}}
\hspace{0.1in}
\subfigure[~$p=\frac{1}{2}$]{
\label{fig:subfig:b}
\includegraphics[width=1.6in]{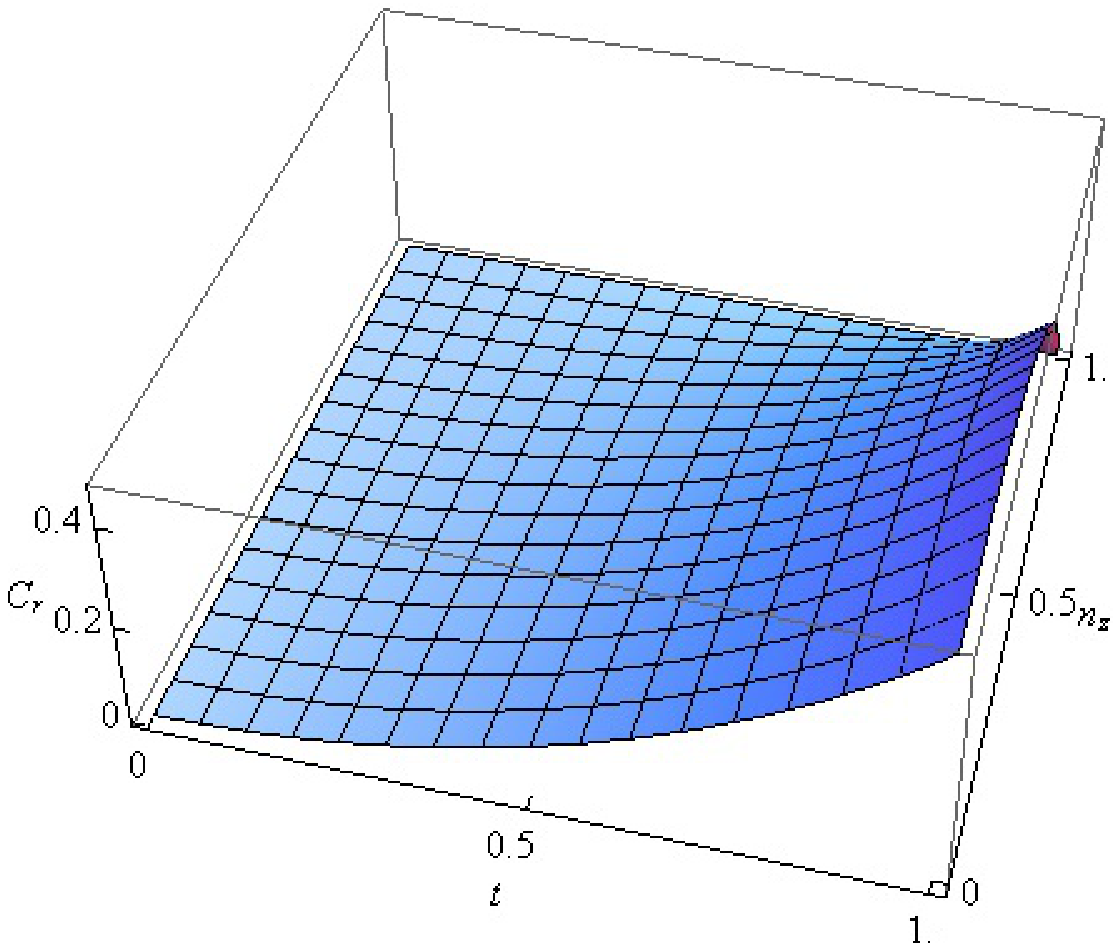}}
\hspace{0.1in}
\subfigure[~$p=\frac{3}{4}$]{
\label{fig:subfig:c}
\includegraphics[width=1.6in]{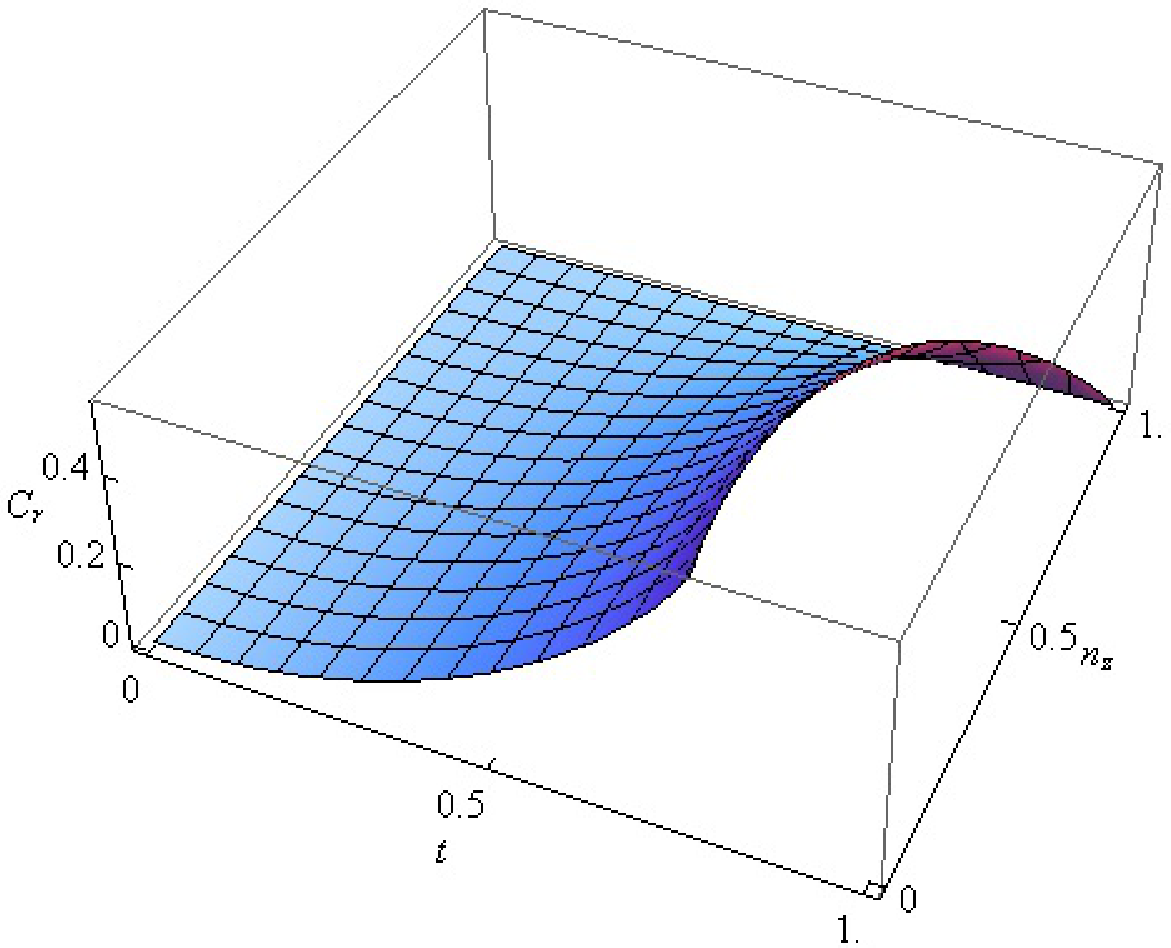}}
\caption{The variation of $\mathcal{C}_r(\varepsilon(\rho))$ with respect to $t$ and $n_z$ under amplitude damping channel.}
\label{fig:subfig1} 
\end{figure}

\begin{figure}
\centering
\subfigure[~$p=\frac{1}{4}$]{
\label{fig:subfig:a}
\includegraphics[width=1.6in]{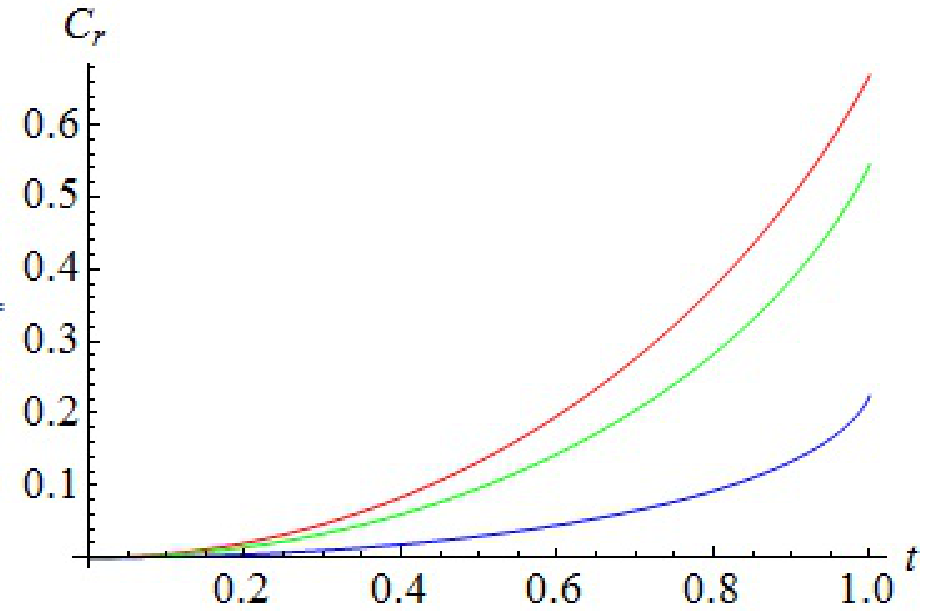}}
\hspace{0.1in}
\subfigure[~$p=\frac{1}{2}$]{
\label{fig:subfig:b}
\includegraphics[width=1.6in]{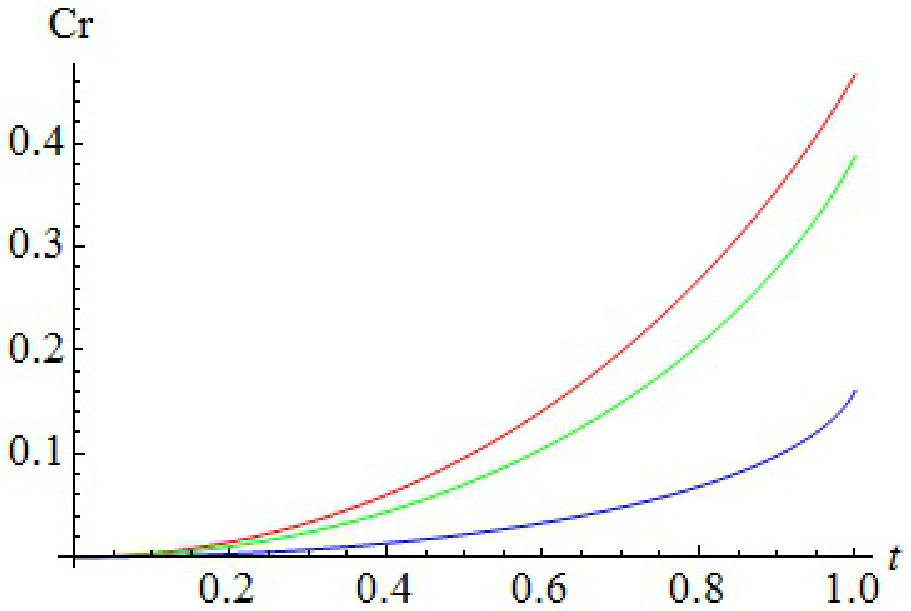}}
\hspace{0.1in}
\subfigure[~$p=\frac{3}{4}$]{
\label{fig:subfig:c}
\includegraphics[width=1.6in]{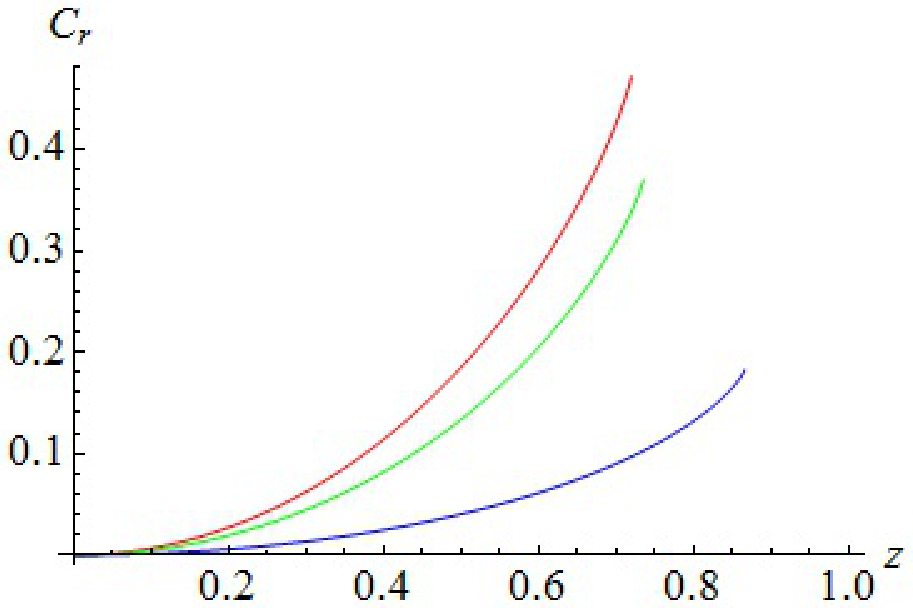}}
\caption{For $p=\frac{1}{4}$, $p=\frac{1}{2}$ and $p=\frac{3}{4}$, $\mathcal{C}_r(\varepsilon(\rho))$
is an increasing function with respect to $t$ for the cases $n_z=0.3$ (red line),
$n_z=0.6$ (green line) and $n_z=0.9$ (blue line).}
\label{fig:subfig2} 
\end{figure}

\begin{figure}
\centering
\subfigure[~$p=\frac{1}{4}$]{
\label{fig:subfig:a}
\includegraphics[width=1.6in]{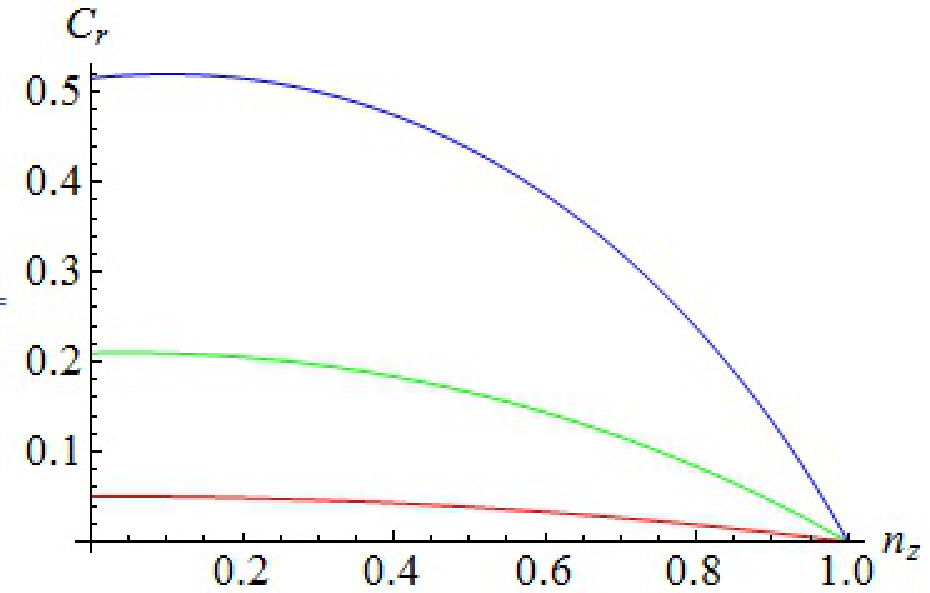}}
\hspace{0.1in}
\subfigure[~$p=\frac{1}{2}$]{
\label{fig:subfig:b}
\includegraphics[width=1.6in]{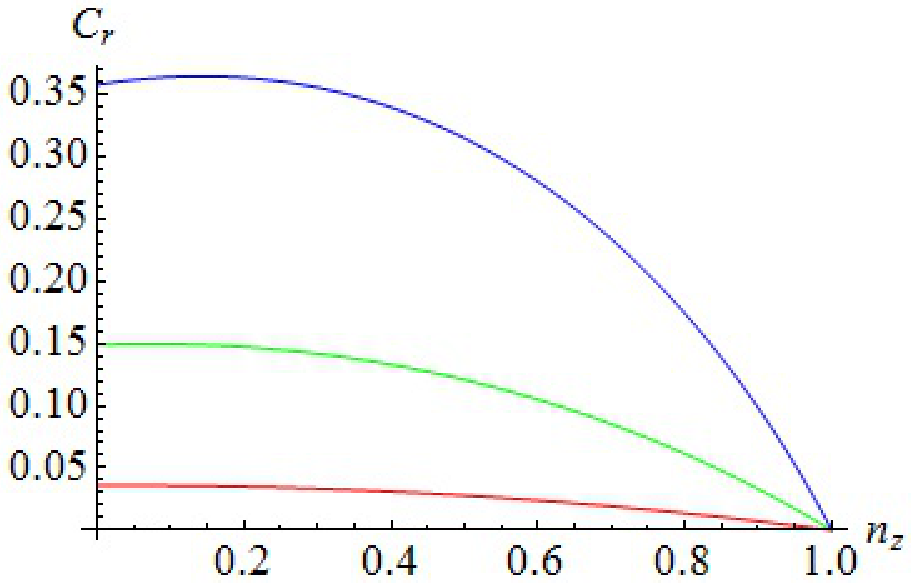}}
\hspace{0.1in}
\subfigure[~$p=\frac{3}{4}$]{
\label{fig:subfig:c}
\includegraphics[width=1.6in]{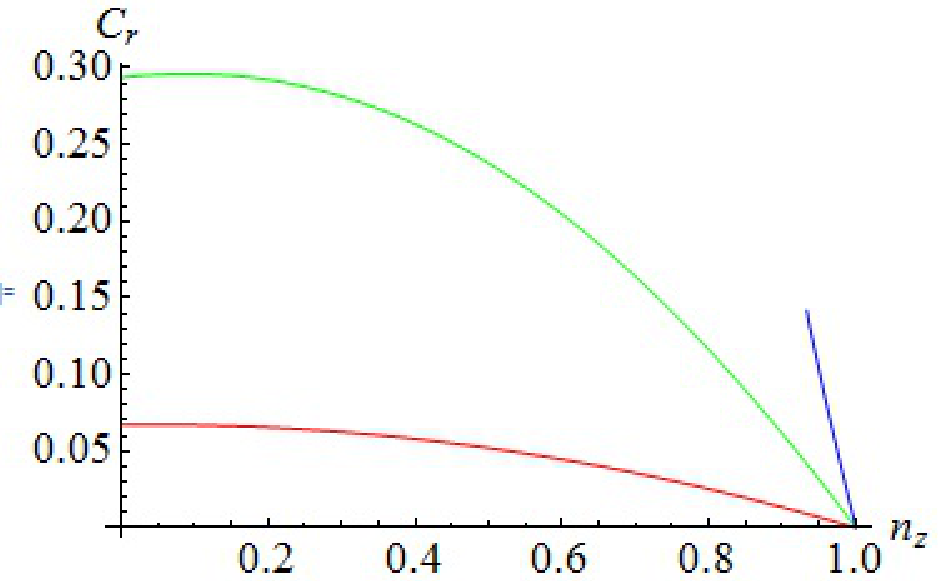}}
\caption{For $p=\frac{1}{4}$, $p=\frac{1}{2}$, and $p=\frac{3}{4}$,
$\mathcal{C}_r(\varepsilon(\rho))$
is a decreasing function with respect to $n_z$ for the cases $t=0.3$ (red line),
$t=0.6$ (green line), and $t=0.9$ (blue line).}
\label{fig:subfig3} 
\end{figure}

It has been proved that the amplitude damping channel does not change the coherence-induced state ordering
under the coherence measure $\mathcal{C}_{l_1}$ \cite{coherent}. In the following
we study the case $p\in(0,1)$ for the coherence measures $\mathcal{C}_r$ and $\mathcal{C}_\alpha$
for $\alpha\in(0,1)\cup(1,2]$.
Numerical calculation shows that for any $p\in(0,1)$, the
amplitude damping channel does not change the coherence-induced state ordering by $\mathcal{C}_r$ with fixed $n_z$ or fixed $t$,
since $\mathcal{C}_r$ is an increasing function with respect to $t$ for every fixed $n_z$ while a decreasing function with respect to
$n_z$ for every fixed $t$; see
figures \ref{fig:subfig1}, \ref{fig:subfig2}, and \ref{fig:subfig3} for the cases of $p=\frac{1}{4}, \ \frac{1}{2}$, and $\frac{3}{4}$.

\begin{figure}
\centering
\subfigure[~$p=\frac{1}{8}$]{
\label{fig:subfig:a}
\includegraphics[width=1.8in]{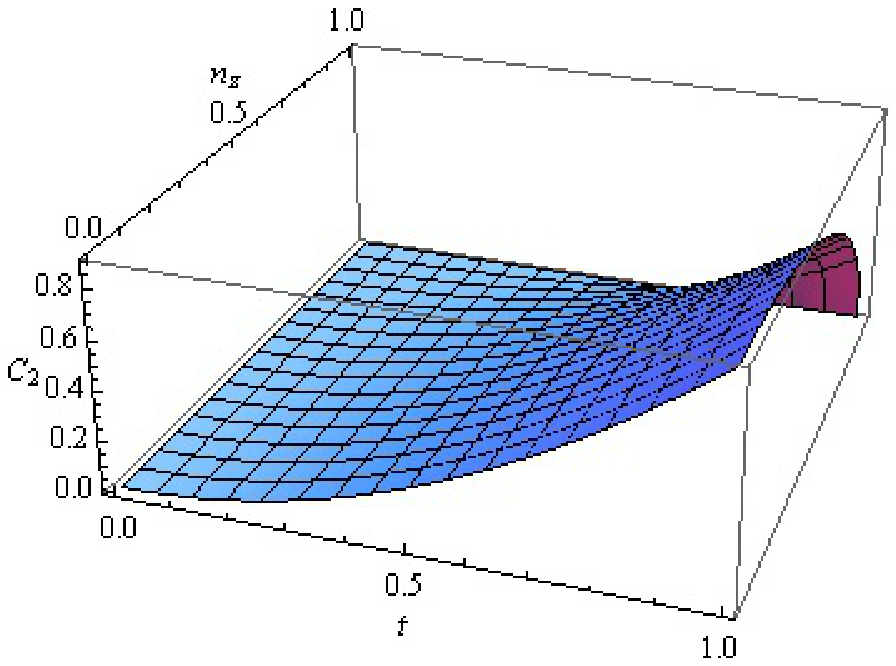}}
\hspace{0.9in}
\subfigure[~$p=\frac{3}{8}$]{
\label{fig:subfig:c}
\includegraphics[width=1.8in]{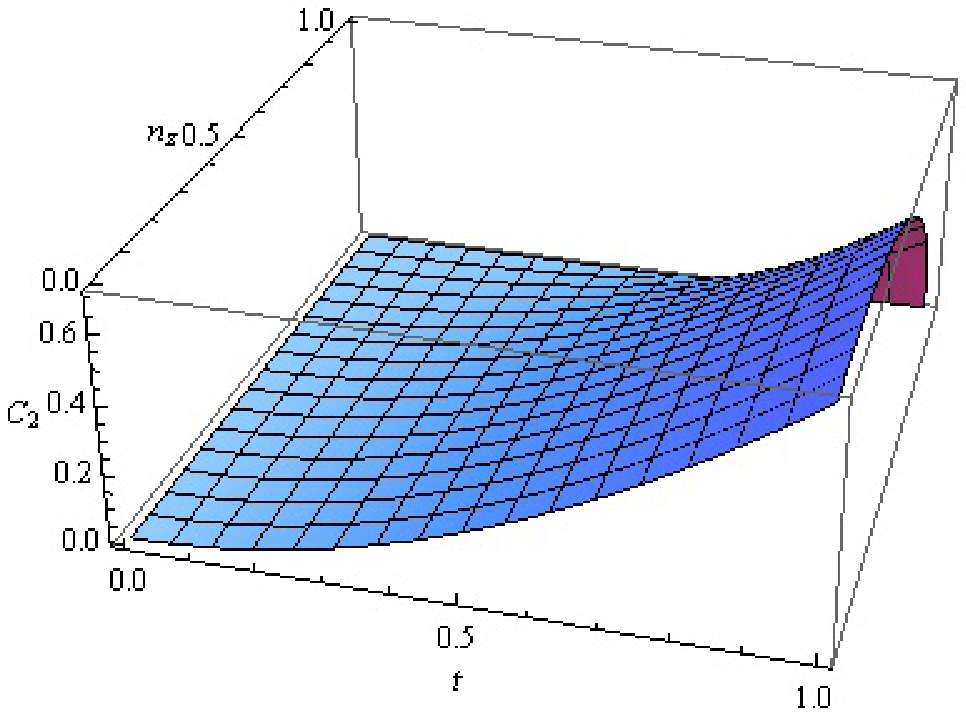}}
\hspace{0.9in}
\subfigure[~$p=\frac{5}{8}$]{
\label{fig:subfig:e}
\includegraphics[width=1.8in]{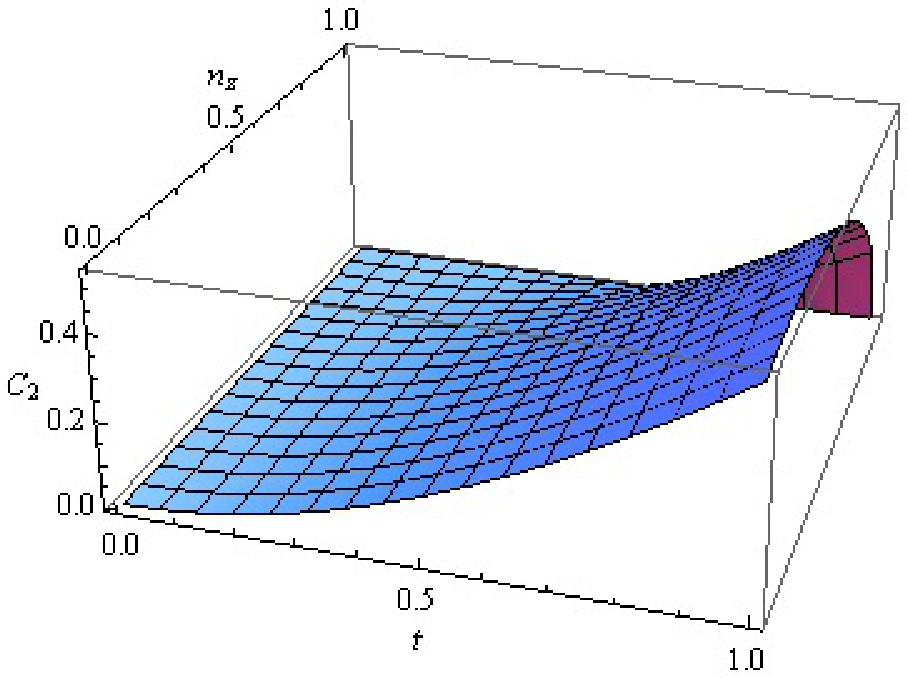}}
\hspace{0.9in}
\subfigure[~$p=\frac{7}{8}$]{
\label{fig:subfig:g}
\includegraphics[width=1.8in]{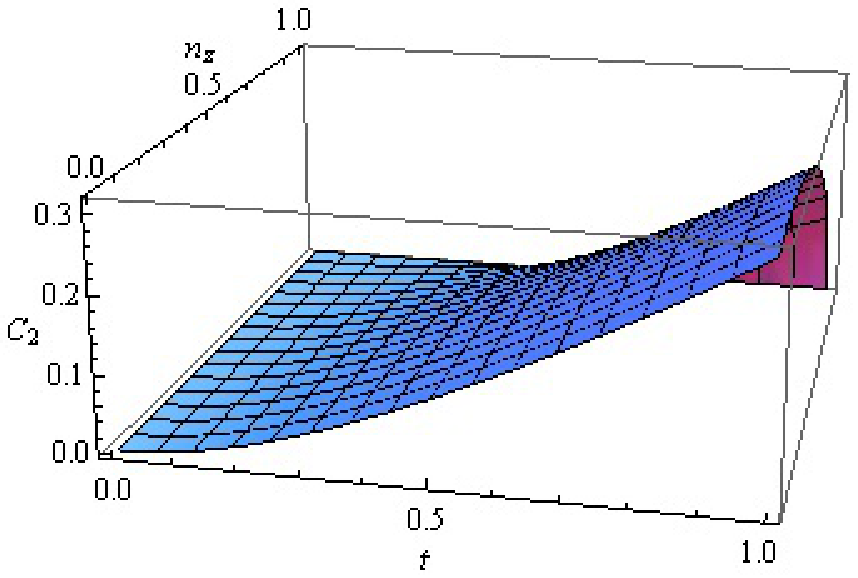}}
\caption{The variation of $\mathcal{C}_2$ with respect to $t$ and $n_z$ under amplitude damping channel for $p=\frac{1}{8}$, $p=\frac{3}{8}$, $p=\frac{5}{8}$,
and $p=\frac{7}{8}$.}
\label{fig:subfig4} 
\end{figure}

\begin{figure}
\centering
\subfigure[~$\alpha=\frac{1}{4}$]{
\label{fig:subfig:a}
\includegraphics[width=1.8in]{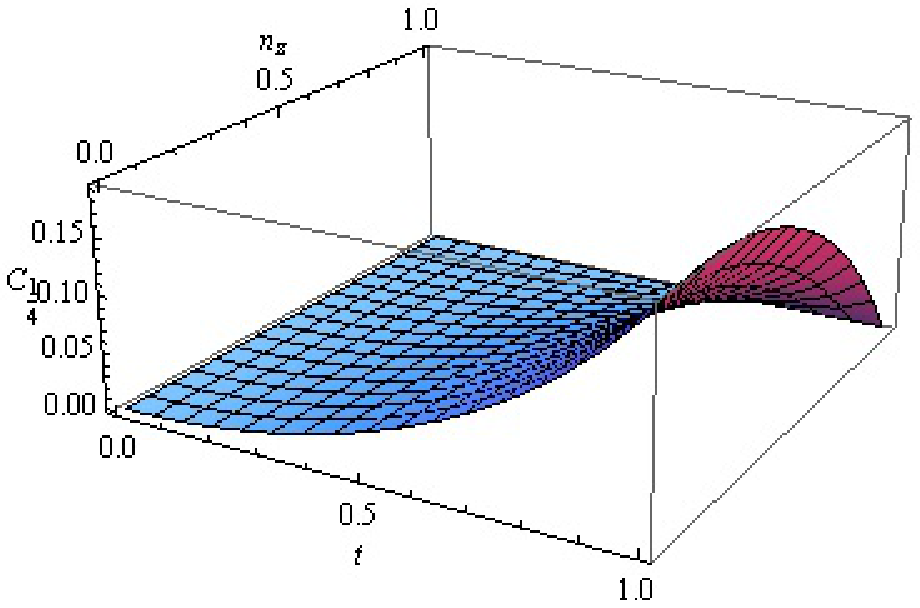}}
\hspace{0.9in}
\subfigure[~$\alpha=\frac{3}{4}$]{
\label{fig:subfig:c}
\includegraphics[width=1.8in]{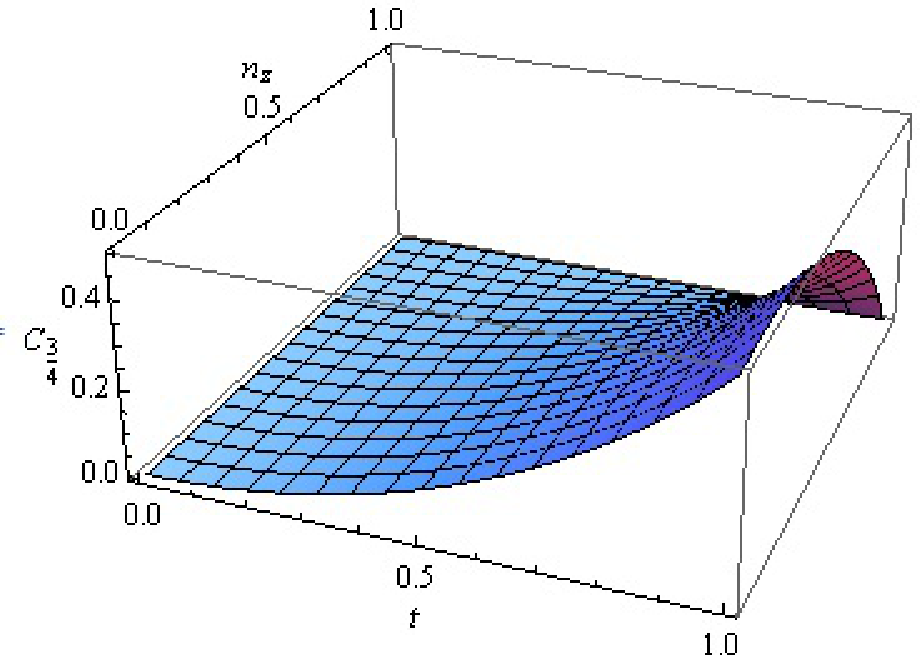}}
\hspace{0.9in}
\subfigure[~$\alpha=\frac{5}{4}$]{
\label{fig:subfig:d}
\includegraphics[width=1.8in]{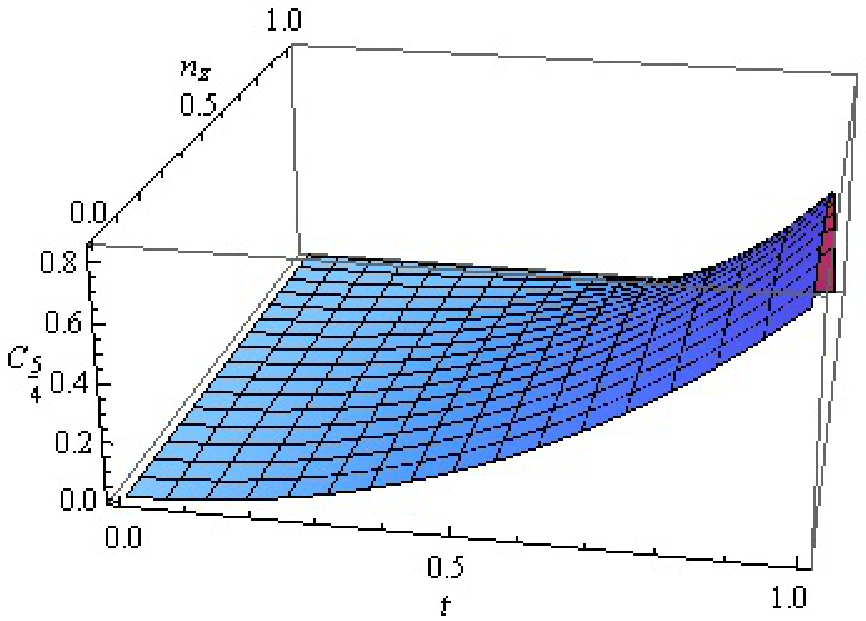}}
\hspace{0.9in}
\subfigure[~$\alpha=\frac{7}{4}$]{
\label{fig:subfig:f}
\includegraphics[width=1.8in]{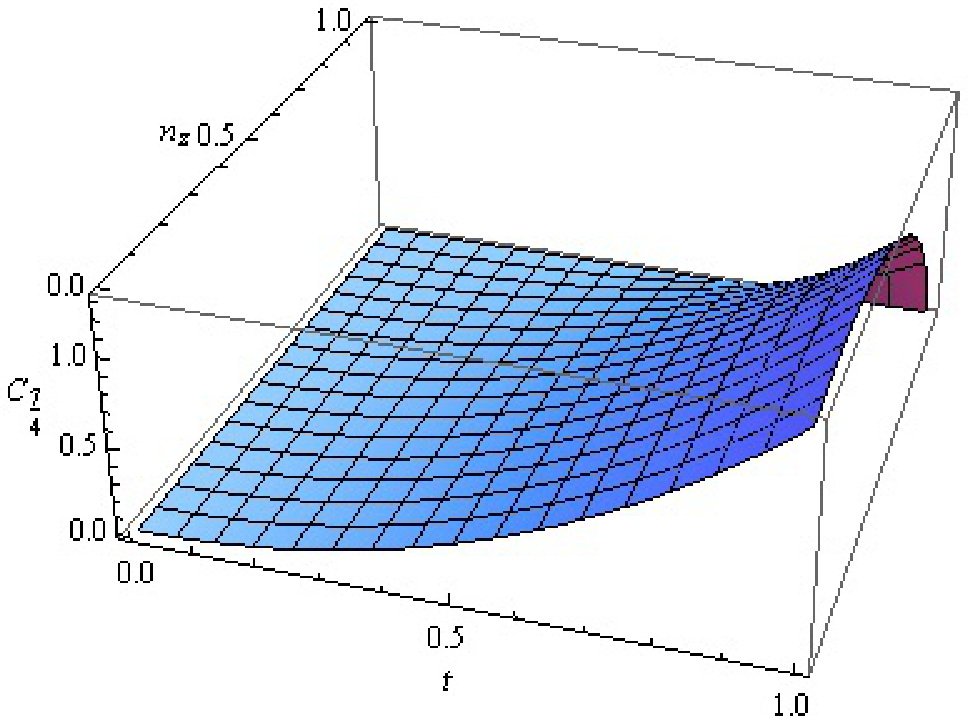}}
\caption{For fixed $p=\frac{1}{2}$, the variation of $\mathcal{C}_\alpha$ with respect to $t$ and $n_z$ under amplitude damping channel for
$\alpha=\frac{1}{4},\,\alpha=\frac{3}{4},\,\alpha=\frac{5}{4}$, and $\alpha=\frac{7}{4}$.}
\label{fig:subfig5} 
\end{figure}

For the coherence measure $\mathcal{C}_{\alpha}$, Zhang {\it et al} have proved that for $p=\frac{1}{2}$,
the amplitude damping channel maintains the coherence-induced state ordering with fixed $n_z$ or fixed $t$.
In fact, we find that it holds for any $p\in (0,1)$ and $\alpha\in(0,1)\cup(1,2]$.
In Fig. \ref{fig:subfig4}, we show the variation of $\mathcal{C}_2$ for $p=\frac{1}{8},\,\frac{3}{8},\,\frac{5}{8}$ and $p=\frac{7}{8}$.
In Fig. \ref{fig:subfig5}, we show the variation of $\mathcal{C}_\alpha$ for fixed $p=\frac{1}{2}$ and
$\alpha=\frac{1}{4},\,\frac{3}{4},\,\frac{5}{4},\,\frac{7}{4}$.

\subsection{Phase damping channel}

Now we study the dynamics of coherence-induced state ordering under phase damping channel,
which can be characterized by the Kraus' operators $K_0=\sqrt{p}I, \ K_1=\sqrt{1-p}|0\rangle\langle0|, \ \ K_2=\sqrt{1-p}|1\rangle\langle1|$,
 where $0\leq p \leq1$.
By applying the phase damping channel to the state represented by Eq. \eqref{sin}, we get
\begin{equation}
\varepsilon(\rho)=\left(
                    \begin{array}{cc}
                      \frac{1+tn_z}{2} & \frac{tp(n_x-\mathbf{i} n_y)}{2} \\
                      \frac{tp(n_x+\mathbf{i} n_y)}{2} & \frac{1-tn_z}{2} \\
                    \end{array}
                  \right).
\end{equation}
For $p=0$, the phase damping channel transforms a state into an incoherent one.
In the following, we study the case $p\neq0$.
For simplicity, we define
$A=1+(p^2-1)(1-n_z)^2$, $B=\frac{1+t\sqrt{A}}{2}$, $C=(\sqrt{A}+n_z)^2$, and $D=p^2(1-n_z^2)$.
Substituting $\varepsilon (\rho)$ into eq. \eqref{l1}, \eqref{rel}, and \eqref{Tsa}, we have
\begin{equation}\label{l_1}
\mathcal{C}_{l_1}(\varepsilon(\rho))=pt\sqrt{1-n_z}=p\mathcal{C}_{l_1}(\rho),
\end{equation}
\begin{equation}\label{Cr}
\mathcal{C}_{r}(\varepsilon(\rho))=h(\frac{1+tn_z}{2})-h(B),
\end{equation}
\begin{equation}\label{eq16}
\mathcal{C}_{\alpha}(\varepsilon(\rho))=\frac{r^\alpha-1}{\alpha-1},
\end{equation}
where $r=(B^\alpha\frac{C}{C+D}+(1-B)^\alpha\frac{D}{C+D})^\frac{1}{\alpha}+((1-B)^\alpha\frac{C}{C+D}+B^\alpha\frac{D}{C+D})^\frac{1}{\alpha}$.
According to Eq. \eqref{l_1} the phase damping channel does not change the coherence-induced state ordering by $\mathcal{C}_{l_1}$
for single-qubit states.

Next we consider the coherence measure $\mathcal{C}_r$.
On differentiating Eq.\eqref{Cr} with respect to $t$, we get
\begin{eqnarray*}
\frac{\partial \mathcal{C}_r(\varepsilon(\rho))}{\partial t}=\frac{n_z}{2}\log \frac{1-tn_z}{1+tn_z}
+\frac{\sqrt{A}}{2}\log\frac{1+t\sqrt{A}}{1-t\sqrt{A}}
\geq \frac{n_z}{2}\log \frac{1-tn_z}{1+tn_z}+\frac{n_z}{2}\log \frac{1+tn_z}{1-tn_z}=0,
\end{eqnarray*}
since $\frac{\sqrt{A}}{2}\log\frac{1+t\sqrt{A}}{1-t\sqrt{A}}$ is an increasing function with respect to $p\in(0,1]$.
Moreover, since $\frac{\partial \mathcal{C}_r(\rho)}{\partial t}\geq 0$,
the phase damping channel does not change the coherence-induced state ordering by $\mathcal{C}_r$
for single-qubit states with fixed $n_z$. On differentiating Eq.\eqref{l_1} with respect to $n_z$, we obtain
$$
\frac{\partial\mathcal{C}_r(\varepsilon(\rho))}{\partial n_z}=(\frac{t}{2}\ln\frac{1-tn_z}{1+tn_z}-\frac{(p^2-1)n_zt}{2\sqrt{A}}
\ln\frac{1+t\sqrt{A}}{1-t\sqrt{A}})/\ln 2.
$$
Set $f(p)=\frac{(p^2-1)}{\sqrt{A}}\ln\frac{1+t\sqrt{A}}{1-t\sqrt{A}}$. Then
\begin{equation*}\begin{array}{rl}
f^{\prime}(p)=\frac{p+pA}{\sqrt{A^3}}\ln\frac{1+t\sqrt{A}}{1-t\sqrt{A}}+\frac{2tp(1-n_z^2)(p^2-1)}{A(1-t^2A)}
\geq \frac{p+pA}{\sqrt{A^3}}\ln\frac{1+t\sqrt{A}}{1-t\sqrt{A}}+\frac{2tp(1-n_z^2)(p^2-1)}{A(1-A)}
=\frac{p}{A}(\frac{A+1}{\sqrt{A}}\ln\frac{1+t\sqrt{A}}{1-t\sqrt{A}}-2t)\geq0,
\end{array}
\end{equation*}
since $\frac{A+1}{\sqrt{A}}\ln\frac{1+t\sqrt{A}}{1-t\sqrt{A}}-2t$ is an increasing function with respect to $t\geq0$.
Thus,
$$
\frac{\partial\mathcal{C}_r(\varepsilon(\rho))}{\partial n_z}\leq(\frac{t}{2}\ln\frac{1-tn_z}{1+tn_z}-\frac{n_z t}{2} f(0))/\ln2=0.
$$
Therefore, the phase damping channel keeps the coherence-induced state ordering by $\mathcal{C}_r$ for single-qubit states
with fixed t as $\frac{\partial\mathcal{C}_r(\rho)}{\partial n_z}\leq0$.

According to Eq.\eqref{Cr}, for the coherence measure $\mathcal{C}_\alpha$, $\alpha\in(0,1)\cup(1,2]$,
we have $\frac{\partial\mathcal{C}_\alpha(\varepsilon (\rho))}{\partial t}
=\frac{\alpha}{\alpha-1}r^{\alpha-1}\frac{\partial r}{\partial t}$,
where
\begin{equation}\begin{array}{rl}
&\frac{\partial r}{\partial t}=
\frac{\sqrt{A}}{2}\{[B^{\alpha}\frac{C}{C+D}+(1-B)^{\alpha}\frac{D}{C+D}]^{\frac{1}{\alpha}-1}[B^{\alpha-1}\frac{C}{C+D}-(1-B)^{\alpha-1}\frac{D}{C+D}]\}\\
&+\frac{\sqrt{A}}{2}\{[(1-B)^{\alpha}\frac{C}{C+D}+B^{\alpha}\frac{D}{C+D}]^{\frac{1}{\alpha}-1}[B^{\alpha-1}\frac{D}{C+D}-(1-B)^{\alpha-1}\frac{C}{C+D}]\}.
\end{array}
\end{equation}

If $\alpha\in (0,1)$,
$\frac{\partial r}{\partial t}\leq[B^{\alpha}\frac{C}{C+D}+(1-B)^{\alpha}\frac{D}{C+D}]^{\frac{1}{\alpha}-1}(B^{\alpha-1}-(1-B)^{\alpha-1})\leq0$.

If $\alpha\in (1,2]$,
$\frac{\partial r}{\partial t}\geq[B^{\alpha}\frac{C}{C+D}+(1-B)^{\alpha}\frac{D}{C+D}]^{\frac{1}{\alpha}-1}(B^{\alpha-1}$ $-(1-B)^{\alpha-1})\geq0$.

\begin{figure}
\centering
\subfigure[~$\alpha=\frac{1}{4}$]{
\label{fig:subfig:a}
\includegraphics[width=1.9in]{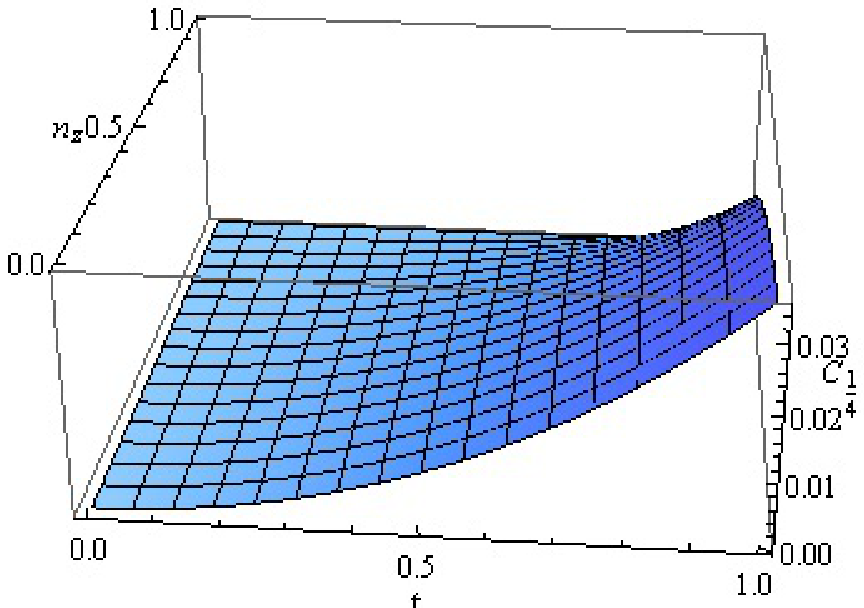}}
\hspace{0.9in}
\subfigure[~$\alpha=\frac{3}{4}$]{
\label{fig:subfig:c}
\includegraphics[width=1.9in]{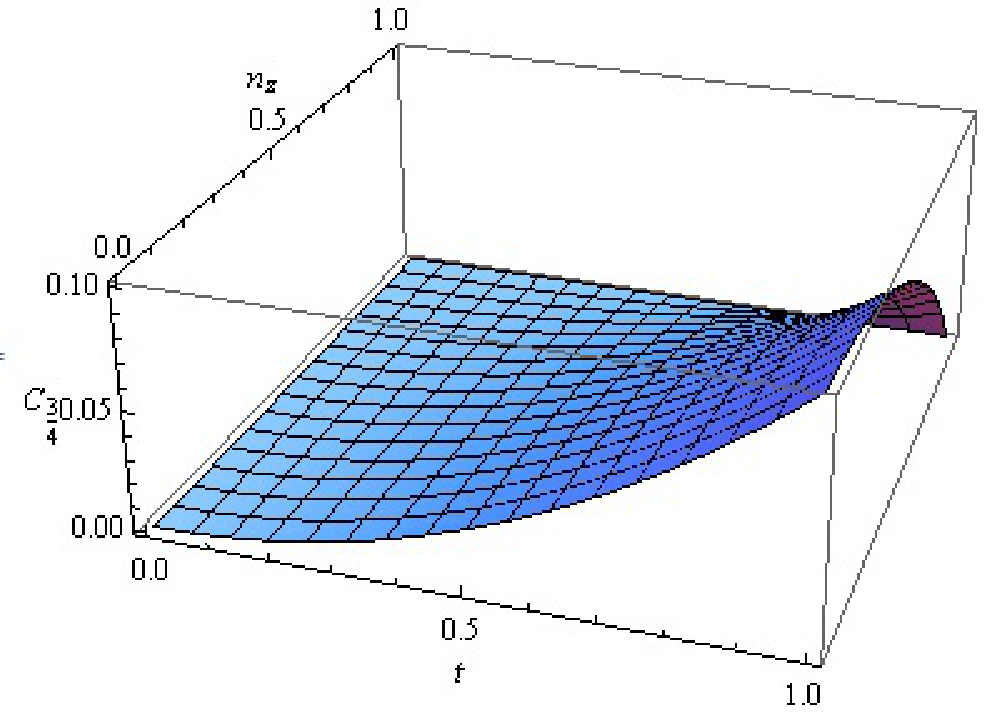}}
\hspace{0.9in}
\subfigure[~$\alpha=\frac{5}{4}$]{
\label{fig:subfig:d}
\includegraphics[width=1.9in]{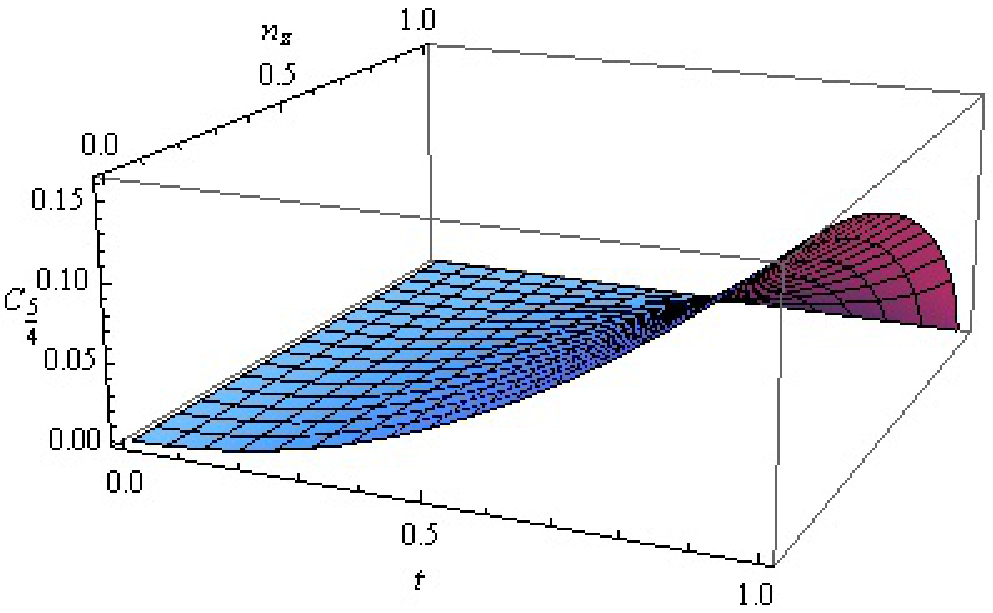}}
\hspace{0.9in}
\subfigure[~$\alpha=\frac{7}{4}$]{
\label{fig:subfig:f}
\includegraphics[width=1.9in]{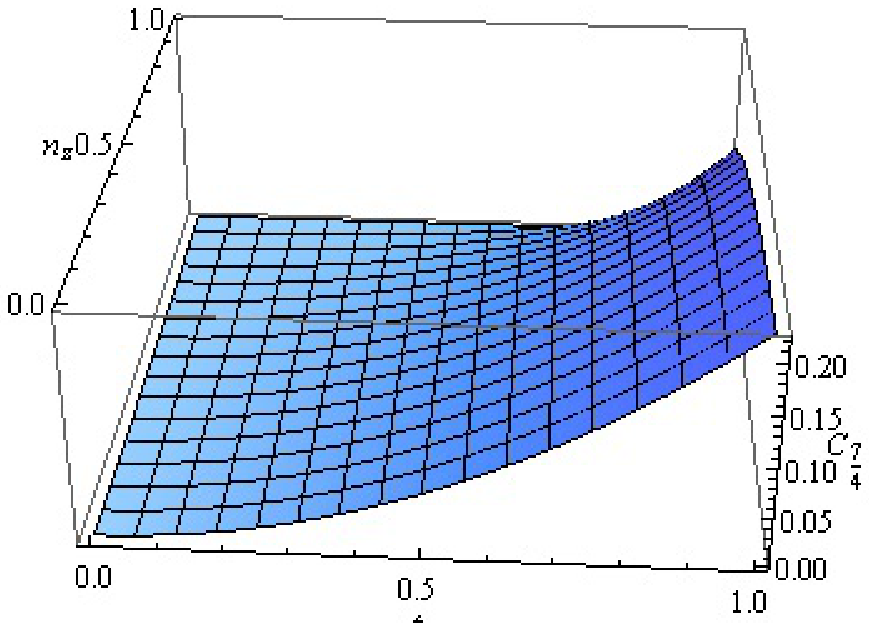}}
\caption{The variation of $\mathcal{C}_\alpha$ with respect to $t$ and $n_z$ under phase damping channel for
fixed $p=\frac{1}{2}$ and $\alpha=\frac{1}{4},\,\frac{3}{4},\,\frac{5}{4}$, and $\alpha=\frac{7}{4}$.}
\label{fig:subfig6} 
\end{figure}

Then $\frac{\partial \mathcal{C}_\alpha(\varepsilon(\rho))}{\partial t}\geq0$.
Since $\frac{\partial \mathcal{C}_\alpha(\rho)}{\partial t}\geq0$, the phase damping channel does not change the coherence-induced state ordering by $\mathcal{C}_\alpha$ for
single-qubit states with fixed $n_z$.
In fact, the phase damping channel does not change the coherence-induced state ordering by $\mathcal{C}_\alpha$ for
single-qubit states with fixed $t$.
In general, it is very difficult to discuss the monotony of $\mathcal{C}_{\alpha}$ for all parameters
$\alpha\in(0,1)\cup(1,2]$ and $p\in (0,1]$ with respect to $n_z$. In
Fig. \ref{fig:subfig6} we present the variation of $\mathcal{C}_\alpha$ with fixed $p=\frac{1}{2}$ for
$\alpha=\frac{1}{4},\,\frac{3}{4},\,\frac{5}{4}$, and $\alpha=\frac{7}{4}$.

\subsection{Depolarizing channel}
Now we study the dynamics of coherence-induced state ordering under depolarizing channel.
The state of the quantum system after depolarizing channel is given by $\varepsilon(\rho)=\frac{pI}{2}+(1-p)\rho$,
\begin{equation}
\varepsilon(\rho)=\left(
                    \begin{array}{cc}
                      \frac{1+tn_z(1-p)}{2} & \frac{(1-p)(n_x-\mathbf{i} n_y)t}{2} \\
                      \frac{(1-p)(n_x+\mathbf{i} n_y)t}{2} & \frac{1-tn_z(1-p)}{2} \\
                    \end{array}
                  \right).
\end{equation}
Substituting $\varepsilon (\rho)$ into eq. \eqref{l1}, \eqref{rel}, and \eqref{Tsa}, we have
\begin{equation}\label{L1}
\mathcal{C}_{l_1}(\varepsilon(\rho))=(1-p)t\sqrt{1-n_z}=(1-p)\mathcal{C}_{l_1}(\rho),
\end{equation}
\begin{equation}\label{eq12}
\mathcal{C}_{r}(\varepsilon(\rho))=h(\frac{1+tn_z(1-p)}{2})-h(\frac{1+t(1-p)}{2}),
\end{equation}
\begin{equation}\label{eq13}
\mathcal{C}_{\alpha}(\varepsilon(\rho))=\frac{r^\alpha-1}{\alpha-1},
\end{equation}
where
$r=[E^\alpha F+(1-E)^\alpha(1-F)]^{\frac{1}{\alpha}}+[E^\alpha(1-F)+(1-E)^\alpha F]^{\frac{1}{\alpha}}$, and
$E=\frac{1+t(1-p)}{2}$, $F=\frac{1+n_z}{2}$.


According to Eq. \eqref{L1}, the depolarizing channel keeps the coherence-induced state ordering under $\mathcal{C}_{l_1}$ for single-qubit states.

Next, we consider the coherence measure $\mathcal{C}_r$.
$\mathcal{C}_r(\varepsilon(\rho))$ is clearly a decreasing function with respect to $n_z$,
since $\frac{\partial \mathcal{C}_r(\varepsilon(\rho))}{\partial n_z}=\frac{t(1-p)}{2}\log \frac{1-tn_z(1-p)}{1+tn_z(1-p)}\leq 0$.
Thus,
the depolarizing channel does not change the coherence-induced state ordering by $\mathcal{C}_r$ for single-qubit states with fixed $t$,
due to the fact that $\mathcal{C}_r(\rho)$ is also a decreasing function with respect to $n_z$.
In fact, $\mathcal{C}_r(\varepsilon(\rho))$ is an increasing function with respect to $t$,
as
$$
\frac{\partial \mathcal{C}_r(\varepsilon(\rho))}{\partial t\partial n_z}=\frac{(1-p)}{2}\log \frac{1-tn_z(1-p)}{1+tn_z(1-p)}-
\frac{t(1-p)^2n_z}{1-t^2n_z^2(1-p)^2}\leq 0.
$$
Therefore,
$$
\frac{\partial \mathcal{C}_r(\varepsilon(\rho))}{\partial t}=\frac{(1-p)n_z}{2}\log \frac{1-tn_z(1-p)}{1+tn_z(1-p)}+
\frac{1-p}{2}\log \frac{1+t(1-p)}{1-t(1-p)}\geq 0.
$$
In addition, $\mathcal{C}_r(\rho)$ is also an increasing function with respect to $t$ \cite{coherent}.
Thus, the depolarizing channel does not change the coherence-induced state ordering by $\mathcal{C}_r$ for single-qubit states with fixed $n_z$.

In the end, we consider the coherence measure $\mathcal{C}_\alpha$, where $\alpha\in(0,1)\cup(1,2]$.
First of all, we show
$\frac{\partial r}{\partial t}\geq 0$ if $\alpha\in(1,2]$ and $\frac{\partial r}{\partial t}\leq 0$ if $\alpha\in(0,1)$.
We have
\begin{equation}
\begin{array}{rl}
&\frac{\partial r}{\partial t}
=\frac{1-p}{2}\{[E^\alpha F+(1-E)^\alpha(1-F)]^{\frac{1}{\alpha}-1}[E^{\alpha-1}F-(1-E)^{\alpha-1}(1-F)]+\\
&[E^\alpha(1-F)+(1-E)^\alpha F]^{\frac{1}{\alpha}-1}[E^{\alpha-1}(1-F)-(1-E)^{\alpha-1} F]\}.
\end{array}\end{equation}
Since $x^{\alpha}y+(1-x)^{\alpha}(1-y)\geq x^{\alpha}(1-y)+(1-x)^{\alpha}y$, where $\alpha>0,\frac{1}{2}\leq x,y\leq 1$,
we have
\begin{equation}
\frac{\partial r}{\partial t}\geq\frac{1-p}{2}[E^\alpha F+(1-E)^\alpha(1-F)]^{\frac{1}{\alpha}-1}[E^{\alpha-1}-(1-E)^{\alpha-1}]\geq0
\end{equation}
if $\alpha\in(1,2]$.
If $\alpha\in(0,1)$, we have
\begin{equation}
\frac{\partial r}{\partial t}\leq\frac{1-p}{2}[E^\alpha F+(1-E)^\alpha(1-F)]^{\frac{1}{\alpha}-1}[E^{\alpha-1}-(1-E)^{\alpha-1}]\leq0.
\end{equation}


Thus, $\frac{\partial \mathcal{C}_\alpha(\varepsilon(\rho))}{\partial t}=\frac{\alpha r^{\alpha-1}}{\alpha-1}\frac{\partial r}{\partial t}\geq0$.
Since $\frac{\partial \mathcal{C}_\alpha(\rho)}{\partial t}\geq 0$,
we arrive at the conclusion that the depolarizing channel does not change the coherence-induced state ordering by $\mathcal{C}_\alpha$ for single-qubit states with fixed $n_z$.

On the other hand, as
$$
\frac{\partial r}{\partial n_z}
=\frac{1}{2\alpha}[E^\alpha-(1-E)^\alpha]\{[E^{\alpha-1}F-(1-E)^{\frac{1}{\alpha}-1}(1-F)]^{\frac{1}{\alpha}-1}-
[E^\alpha(1-F)+(1-E)^\alpha F]^{\frac{1}{\alpha}-1}\}
$$
and $x^{\alpha}y-(1-x)^\alpha(1-y)\geq x^{\alpha}(1-y)-(1-x)^\alpha y$ for $\alpha\geq0$ and $\frac{1}{2}\leq x,y\leq 1$,
one has $\frac{\partial \mathcal{C}_\alpha(\varepsilon(\rho))}{\partial n_z}\leq 0$.
Also, $\frac{\partial \mathcal{C}_\alpha(\rho)}{\partial n_z}\leq 0$, therefore,
we conclude that the depolarizing channel keeps the coherence-induced state ordering by $\mathcal{C}_\alpha$
for single-qubit states with fixed $t$.

\subsection{Bit flit channel}
Now we study the dynamics of coherence-induced state ordering under bit flit channel,
which can be characterized by the Kraus' operators $K_0=\sqrt{p}I, \ K_1=\sqrt{1-p}\sigma_x$, where $0\leq p \leq1$.
Applying the bit flit channel to the state represented by Eq. \eqref{sin}, we get
\begin{equation}
\varepsilon(\rho)=\left(
                    \begin{array}{cc}
                      \frac{1+tn_z(2p-1)}{2} & \frac{tn_x-\mathbf{i} tn_y(2p-1)}{2} \\
                      \frac{tn_x+\mathbf{i} tn_y(2p-1)}{2} & \frac{1-tn_z(2p-1)}{2} \\
                    \end{array}
                  \right).
\end{equation}
Substituting this $\varepsilon (\rho)$ into Eqs. \eqref{l1}, \eqref{rel}, and \eqref{Tsa}, we have
\begin{equation}\label{eq17}
\mathcal{C}_{l_1}(\varepsilon(\rho))=\sqrt{t^2n_x^2+(2p-1)^2t^2n_y^2}=\sqrt{(2p-1)^2\mathcal{C}_{l_1}^2(\rho)+4(p-p^2)t^2n_x^2},
\end{equation}
\begin{equation}\label{eq18}
\mathcal{C}_{r}(\varepsilon(\rho))=h(\frac{1+tn_z(2p-1)}{2})-h(H),
\end{equation}
\begin{equation}\label{eq19}
\mathcal{C}_{\alpha}(\varepsilon(\rho))=\frac{r^\alpha-1}{\alpha-1},
\end{equation}
where
$r=[H^\alpha\frac{M}{M+N}+(1-H)^\alpha\frac{N}{M+N}]^{\frac{1}{\alpha}}+[(1-H)^\alpha\frac{M}{M+N}+
H^\alpha\frac{N}{M+N}]^{\frac{1}{\alpha}}$, and
$G=
\sqrt{1+4(p^2-p)(1-n_x^2)}$, $H=\frac{1+t\sqrt{G}}{2}$, $M=n_x^2+(2p-1)^2n_y^2$ and $N=(\sqrt{G}-(2p-1)n_z)^2$.

Let us consider the special case $p=\frac{1}{2}$. Thus,
\begin{equation}\label{eq20}
\mathcal{C}_{l_1}(\varepsilon(\rho))=tn_x,
\end{equation}
\begin{equation}\label{eq18}
\mathcal{C}_{r}(\varepsilon(\rho))=1-h(\frac{1+tn_x}{2}),
\end{equation}
\begin{equation}\label{eq19}
\mathcal{C}_{\alpha}(\varepsilon(\rho))=\frac{r^\alpha-1}{\alpha-1},
\end{equation}
where $r=2[\frac{1}{2}(\frac{1+tn_x}{2})^\alpha+\frac{1}{2}(\frac{1-tn_x}{2})^\alpha]^{\frac{1}{\alpha}}$.
Hence, $\frac{\partial \mathcal{C}_{l_1}(\varepsilon(\rho))}{\partial n_x}=t\geq0$,
$\frac{\partial \mathcal{C}_{r}(\varepsilon(\rho))}{\partial n_x}=\frac{t}{2}\log \frac{1+tn_x}{1-tn_x}\geq0$
and
$$
\frac{\partial \mathcal{C}_{\alpha}(\varepsilon(\rho))}{\partial n_x}=
\frac{\alpha t}{2(\alpha-1)}r^{\alpha-1}[\frac{1}{2}(\frac{1+tn_x}{2})^\alpha+\frac{1}{2}(\frac{1-tn_x}{2})^\alpha]^{\frac{1}{\alpha}-1}
[(\frac{1+tn_x}{2})^{\alpha-1}-(\frac{1-tn_x}{2})^{\alpha-1}]\geq0.
$$

Let $\rho_1=\frac{1}{2}(I+t_1\vec{n_1}\vec{\sigma})$ and $\rho_2=\frac{1}{2}(I+t_2\vec{n_2}\vec{\sigma})$,
where $n_1=({n_1}_x,{n_1}_y,{n_1}_z)$, $n_2=({n_2}_x,{n_2}_y,{n_2}_z)$.
Assume $t_1=t_2,\ {n_1}_x<{n_2}_x$ and ${n_1}_z<{n_2}_z$.
Then we find that $\mathcal{C}_{l_1}(\rho_1)>\mathcal{C}_{l_1}(\rho_2),\mathcal{C}_{l_1}(\varepsilon(\rho_1))<\mathcal{C}_{l_1}(\varepsilon(\rho_2))$,
$\mathcal{C}_{r}(\rho_1)>\mathcal{C}_{r}(\rho_2),\mathcal{C}_{r}(\varepsilon(\rho_1))<\mathcal{C}_{r}(\varepsilon(\rho_2))$,
$\mathcal{C}_{\alpha}(\rho_1)>\mathcal{C}_{\alpha}(\rho_2)$, and $\mathcal{C}_{\alpha}(\varepsilon(\rho_1))<\mathcal{C}_{\alpha}(\varepsilon(\rho_2))$.
Thus, the bit flit channel changes the coherence-induced state ordering by the coherence measures $\mathcal{C}_{l_1},\ \mathcal{C}_{r}$,
and $\mathcal{C}_{\alpha}$ for single-qubit states with fixed $t$, where $\alpha\in(0,1)\cup(1,2]$.

Now assume $t_1>t_2,\ {n_1}_x<{n_2}_x$ and ${n_1}_z={n_2}_z$ such that $t_1{n_1}_x<t_2{n_2}_x$.
Then we find that $\mathcal{C}_{l_1}(\rho_1)>\mathcal{C}_{l_1}(\rho_2),\mathcal{C}_{l_1}(\varepsilon(\rho_1))<\mathcal{C}_{l_1}(\varepsilon(\rho_2))$,
$\mathcal{C}_{r}(\rho_1)>\mathcal{C}_{r}(\rho_2),\mathcal{C}_{r}(\varepsilon(\rho_1))<\mathcal{C}_{r}(\varepsilon(\rho_2))$,
$\mathcal{C}_{\alpha}(\rho_1)>\mathcal{C}_{r}(\rho_2)$ and $\mathcal{C}_{\alpha}(\varepsilon(\rho_1))<\mathcal{C}_{r}(\varepsilon(\rho_2))$,
since the coherence measures $\mathcal{C}_{l_1}, \ \mathcal{C}_{r}$ and $\mathcal{C}_{\alpha}$ are all increasing functions with respect to $tn_x$.
Thus, bit flit channel changes the coherence-induced state ordering by the coherence measures $\mathcal{C}_{l_1},\ \mathcal{C}_{r}$
and $\mathcal{C}_{\alpha}$ for single-qubit states with fixed $n_z$, where $\alpha\in(0,1)\cup(1,2]$.

\section{Conclusion}
We have discussed whether or not a quantum channel changes the coherence-induced state ordering,
for four specific Markovian channels: $-$ amplitude damping channel, phase flit channel, depolarizing channel, and bit flit channel.
We have shown that the depolarizing channel
does not change the coherence-induced state ordering by $\mathcal{C}_{l_1}$, $\mathcal{C}_r$, $\mathcal{C}_{\alpha}$, and $\mathcal{C}_g$.
For the bit flit channel, we have shown that it does change the coherence-induced state ordering under these four coherence measures
for the case of $p=\frac{1}{2}$. Our results enrich the understanding of coherence-induced state ordering
under quantum channels.


\begin{thebibliography}{00}
\bibitem{berg1} {\AA}berg, J.: Catalytic coherence. Phys. Rev. Lett. $\mathbf{113}$, 150402 (2014).
\bibitem{nara} Narasimhachar, V., Gour, G.: Low-temperature thermodynamics with quantum coherence. Nat. Commun. $\mathbf{6}$, 7689 (2015).
\bibitem{cwik} \'{C}wikli\'{n}ski, P., Studzinski, M., Horodecki, M., Oppenheim, J.: Limitations on the evolution of quantum coherences:
towards fully quantum second laws of thermodynamics. Phys. Rev. Lett. $\mathbf{115}$, 210403 (2015).
\bibitem{lost1} Lostaglio, M., Jennings, D., Rudolph, T.: Description of quantum coherence in thermodynamic processes requires constraints
 beyond free energy. Nat. Commun. $\mathbf{6}$, 6383 (2015).
\bibitem{lost2} Lostaglio, M., Korzekwa, K., Jennings, D., Rudolph, T.: Quantum coherence, timetranslation symmetry, and thermodynamics.
Phys. Rev. X $\mathbf{5}$, 021001 (2015).
\bibitem{plenio} Plenio, M. B., Huelga, S. F.: Dephasing-assisted transport: quantum networks and biomolecules. New J. Phys. $\mathbf{10}$, 113019 (2008).
\bibitem{reben} Rebentrost, P., Mohseni, M., Aspuru-Guzik, A.: Role of quantum coherence and environmental fluctuations in chromophoric energy transport.
J. Phys. Chem. B $\mathbf{113}$, 9942 (2009).
\bibitem{lloyd} Lloyd, S.: Quantum coherence in biological systems. J. Phys. Conf. Ser. $\mathbf{302}$, 012037 (2011).
\bibitem{li} Li, C.-M., Lambert, N., Chen, Y.-N., Chen, G.-Y., Nori, F.: Witnessing quantum coherence: from solid-state to biological systems.
Sci. Rep. $\mathbf{2}$, 885 (2012).
\bibitem{huel} Huelga, S., Plenio, M.: Vibrations, quanta and biology. Contemp. Phys. $\mathbf{54}$, 181 (2013).
\bibitem{levi} Levi, F., Mintert, F.: A quantitative theory of coherent delocalization. New J. Phys. $\mathbf{16}$, 033007 (2014).
\bibitem{vazq} Vazquez, H., Skouta, R., Schneebeli, S., Kamenetska, M., Breslow, R., Venkataraman, L., Hybertsen, M.:
Probing the conductance superposition law in singlemolecule circuits with parallel paths. Nat. Nanotechnol. $\mathbf{7}$, 663 (2012)
\bibitem{karl} Karlstr\"{o}m, O., Linke, H., Karlstrom, G., Wacker, A.: Increasing thermoelectric performance using coherent transport.
Phys. Rev. B $\mathbf{84}$, 113415 (2011).
\bibitem{berg2} J. {\AA} berg.: Quanatifying superposition. arXiv:0612146.
\bibitem{baum} Baumgratz, T., Cramer, M., Plenio, M. B.: Quantifying coherence. Phys. Rev. Lett. $\mathbf{113}$, 140401 (2014).
\bibitem{stre} Streltsov, A., Singh, U., Dhar, H. S., Bera, M. N., Adesso, G.: Measuring quantum coherence
with entanglement. Phys. Rev. Lett. $\mathbf{115}$, 020403 (2015).
\bibitem{zhang} Zhang, F.-G., Shao, L.-H., Luo, Y., Li, Y.-M.: Ordering states with Tsallis relative $\alpha$-entropies of coherence.
Quantum Inf Process (2017) 16: 31.
\bibitem{coherent} Zhang., F.-G., Li., Y.-M.: Coherent-induced state ordering with fixed mixedness. arXiv:1704.02244v1.
\bibitem{liu} Liu, C.-L., Yu, X.-D., Xu,G.-F., Tong, D.-M.: Ordering states with coherence measures.
Quantum Inf Process (2016) 15: 4189.
\bibitem{NC} X. Y. Hu, Channels that do not generate coherence. Phys. Rev. A $\mathbf{94}$, 012326 (2016).
\bibitem{ziji} L. M. Yang, B. Chen, S. M. Fei, Z. X. Wang, Ordering states with various coherence measures, Quantum Inf Process (2018) 17: 91.
\end{thebibliography}
\end{document}